\documentstyle[12pt]{article}
\font\bxfnt=cmsy10 scaled\magstep1
\def\BoxOp{\mathord{\hbox{\bxfnt\char116 }\llap{\bxfnt\char117 }}}

\font\titlefont=cmbx10 scaled \magstep3

\begin{document}
\input{epsf}

\begin{flushright}
\vspace*{-2cm}
gr-qc/9608005 \\ 
TUTP-96-3 \\ August 1, 1996
\vspace*{1.5cm}
\end{flushright}

\begin{center}
{\titlefont Quantum Inequalities on the Energy Density\\
\vspace*{0.1in}
in Static Robertson-Walker Spacetimes\\}
\vskip .7in
\baselineskip=24pt
Michael J. Pfenning\footnote{email: mitchel@tuhepa.phy.tufts.edu} and 
        L.H. Ford\footnote{email: ford@cosmos2.phy.tufts.edu}\\
\vspace*{0.3in}
        Institute of Cosmology \\
        Department of Physics and Astronomy \\ 
        Tufts University \\
        Medford, Massachusetts 02155
\end{center}
\vskip 0.3in

\begin{abstract}
Quantum inequality restrictions on the stress-energy
tensor for negative energy are developed for three and four-dimensional
static spacetimes. We derive a general inequality in terms of a sum of
mode functions which constrains the magnitude and duration of negative
energy seen by an observer at rest in a static spacetime. This inequality
is evaluated explicitly for a minimally coupled scalar field in three
and four-dimensional static Robertson-Walker universes. In the limit of
vanishing curvature, the flat spacetime inequalities are recovered. More
generally, these inequalities contain the effects of spacetime curvature.
In the limit of short sampling times, they take the flat space form plus
subdominant curvature-dependent corrections.
\end{abstract}
\newpage

\baselineskip=14pt

\section{Introduction}

Recently, several approaches have been proposed to study the extent
to which quantum fields may violate the ``weak energy condition,''
(WEC) $T_{\mu\nu}u^\mu u^\nu \geq 0$, for all timelike vectors $u^\mu$.
Such violations occur in the Casimir effect and in quantum coherence
effects, where the energy density in a region may be negative.  
Violations of the local energy conditions
first led to the averaging of the energy condition over timelike
(or null) geodesics\cite{awec}. This eventually led to the ``averaged
weak energy condition'',
\begin{equation}
\int_{-\infty}^\infty T_{\mu\nu}u^\mu u^\nu  d\tau \geq 0. \label{eq:awec}
\end{equation}
Here the total energy density averaged along an entire geodesic
is constrained to be positive. While this does preserve in
some form the weak energy condition, it does not constrain the
magnitude of the negative energy violations from becoming arbitrarily
large over an arbitrary interval, so long as there is compensating
positive energy elsewhere. In cases such as the Casimir effect, where there
is a constant negative energy density present in the vacuum state, 
Eq.~(\ref{eq:awec}) fails. However, it is often still possible to
prove ``difference inequalities'' in which $T_{\mu\nu}$ is replaced by the
difference in expectation values between an arbitrary state and the 
vacuum state \cite{F&Ro95,Yurtsever}. 

Quantum inequality (QI) type relations have been proven 
\cite{F&Ro95,ford78,F&Ro92} which do constrain the magnitude and extent of
negative energy. For example, in  4D Minkowski space
the quantum inequality for free, quantized, massless scalar fields
can be written in its covariant form as
\begin{equation}
\hat\rho = {\tau_0 \over \pi} \int_{-\infty}^\infty 
{\langle T_{\mu\nu}u^\mu u^\nu \rangle d\tau \over {\tau^2 + \tau_0^2}}
\geq -{3\over 32\pi^2 \tau_0^4},   \label{eq:qi1}
\end{equation}
for all values of $\tau_0$ and where $u^\mu$ is a timelike vector.
The expectation value $\langle \rangle$ is with respect to an
arbitrary state $|\psi\rangle$, and $\tau_0$ is the characteristic
width of the sampling function, $\tau_0 /\pi (\tau^2 + \tau_0^2)$,
whose time integral is unity.  Such inequalities limit the
magnitude of the negative energy violations and the time for
which they are allowed to exist. In addition, such
QI relations reduce to the usual AWEC type conditions in the infinite
sampling time limit.

Flat space quantum inequality-type relations of this form have 
since been applied to curved spacetimes \cite{F&Ro96,Mitch} by
keeping the sampling time shorter than a ``characteristic'' curvature
radius of the geometry.  Under such circumstances, it was argued
that the spacetime is approximately flat, and the inequalities
could be applied over a small region. In the wormhole geometry
\cite{F&Ro96}, this led to wormholes which were either on the order
of a Planck length in size or with a great disparity in the
length scales that characterize the wormhole.  In the ``warp drive''
geometry \cite{Alcu94} it was found that if one wanted superluminal
bubbles of macroscopically useful size then the bubble wall thickness
would be of the order of a Planck length \cite{Mitch}. Although the
method of small sampling times is useful, it does not address the
question of how the curvature would enter into the quantum inequalities
for arbitrarily long sampling times.  

In this paper we will address this issue.  We will begin with
a generalized theory in Section 2 which will allow us to find quantum 
difference inequalities in a globally static, but arbitrarily curved spacetime
Then in Section 3 we will look at the case of the $S^2 \times R$
spacetime, the 3D equivalent to the Einstein universe.  We will show
that the QI's are modified by a ``scale function'' which is dependent
upon the ratio of sampling time to the curvature radius ($t_0/a$).
In Section 4, we will proceed to find similar QI's for the
three cases of the 4D static Robertson-Walker spacetimes.  In conclusion,
we will discuss how AWEC type conditions in these particular
spacetimes can be found.  We will use units in which $\hbar = c =
G = 1$.

\section{General Theory}

In this section, we will develop a formalism to find the quantum inequalities 
for a massive, minimally coupled scalar field in a generalized spacetime.  
This method will be
applicable in globally static spacetimes, allowing us to use a
separation of variables of the wave equation, and write the positive
frequency mode functions as
\begin{equation}
f_\lambda({\bf x},t) = U_\lambda ({\bf x}) e^{-i\omega t}.
\end{equation}
The label $\lambda$ represents the set of quantum numbers necessary to
specify the state. Additionally, the mode functions should have unit 
Klein-Gordon norm
\begin{equation}
\left(f_\lambda, f_{\lambda'}\right) = \delta_{\lambda\lambda'}.
\end{equation}

The above separation of variables can be accomplished when 
$\partial_t$ is a timelike Killing vector. Such a spacetime can be
described by a metric of the form
\begin{equation}
{ds}^2 = -{dt}^2 + g_{ij}({\bf x})dx_i dx_j.
\end{equation}
Here $g_{ij}$ is the metric of the spacelike hypersurfaces that are orthogonal
to the Killing vector in the time direction. With this metric, 
the wave equation is
\begin{equation}
\BoxOp\phi - \mu^2 \phi = -\partial_t^2\phi + {1\over \sqrt{|g|}}
\left(\partial_i \sqrt{|g|} g^{ij}\partial_j \phi\right) - \mu^2\phi = 0,
\end{equation}
where $g={\rm det}(g_{\mu\nu})$. The scalar field $\phi$ can then be 
expanded in terms of creation and annihilation operators as
\begin{equation}
\phi = \sum_{\lambda} {\bigl(a_{\lambda}f_{\lambda} + a^\dagger_{\lambda}
f^\ast_{\lambda}\bigr)},
\end{equation}
when quantization is carried out over a finite box or universe.  If 
the spacetime is itself infinite, then we replace the summation by
an integral over all of the possible modes.  The creation and annihilation
operators satisfy the usual commutation relations \cite{Brl&Dv}.

In principle, quantum inequalities can be found for any geodesic 
observer \cite{F&Ro95}. In the  Robertson-Walker universes, static
observers see the universe as having maximal symmetry.  Moving observers
lose this symmetry, and in general the mode functions of the wave equation
can become quite complicated. Thus, for this paper we will concern
ourselves only with static observers, whose four-velocity is the timelike
Killing vector. The energy density that such an
observer samples is given by
\begin{equation}
\rho = T_{\mu\nu} u^\mu u^\nu = T_{00} = {1\over 2}\left[ (\partial_t\phi)^2
+\partial^j\phi \partial_j\phi + \mu^2\phi^2\right]\,. \label{eq:rho}
\end{equation}
Upon substitution of the above mode function expansion into 
Eq.~(\ref{eq:rho}), one finds
\begin{eqnarray}
T_{00} & = & {\rm Re}\sum_{\lambda\lambda'} \left\{{\omega\omega'}
\left[ U_\lambda^* U_{\lambda'} e^{+i(\omega-\omega')t}a_{\lambda}^\dagger
a_{\lambda'}-  U_\lambda U_{\lambda'} e^{-i(\omega+\omega')t}a_{\lambda} 
a_{\lambda'}\right]\right. \nonumber\\
&&\qquad+\left[ \partial^j U_\lambda^* \partial_j 
U_{\lambda'} e^{+i(\omega-\omega')t}a_{\lambda}^\dagger a_{\lambda'}+
\partial^j U_\lambda \partial_j U_{\lambda'} e^{-i(\omega+\omega')t}
a_{\lambda} a_{\lambda'}\right]\nonumber\\
&&\qquad +\mu^2 \left.\left[ U_\lambda^* U_{\lambda'} e^{+i(\omega-\omega')t}
a_\lambda^\dagger a_{\lambda'}+  U_\lambda U_{\lambda'} 
e^{-i(\omega+\omega')t}a_{\lambda} 
a_{\lambda'}\right] \right\} \nonumber\\
&& +{1\over 2} \sum_\lambda \left(\omega^2 U_\lambda^* U_\lambda +
\partial^j U_\lambda^* \partial_j U_\lambda +\mu^2  U_\lambda^* 
U_\lambda\right)
\end{eqnarray}

The last term is the expectation value in the vacuum state, defined by
$a_\lambda |0\rangle = 0$ for all $\lambda$, and is formally divergent.
The vacuum energy density may be defined by a suitable regularization and 
renormalization procedure, as will be discussed in more detail in 
Section~3 and 4. In general, however, it is not uniquely defined.
This ambiguity may be side-stepped by concentrating attention upon
the difference between the energy density in an arbitrary state and that
in the vacuum state, as was done in Ref. \cite{F&Ro95}. 
We will therefore concern ourselves primarily with
\begin{equation}
:T_{00}:\; = T_{00} - \langle 0 |T_{00}| 0\rangle,
\end{equation}
where $| 0\rangle$ represents the Fock vacuum state defined by
the global timelike Killing vector. We will  average the energy density 
as in Eq.~(\ref{eq:qi1}), and find that the averaged energy density difference 
is given by
\begin{eqnarray} 
\Delta\hat \rho &\equiv& {t_0\over\pi}\int_{-\infty}^\infty
{{\langle :T_{00}:\rangle dt} \over{t^2 + t_0^2}}\nonumber\\
& = & {\rm Re}\sum_{\lambda\lambda'}\left\{{\omega\omega'}
\left[ U_\lambda^* U_{\lambda'} e^{-|\omega-\omega'|t_0}\langle 
a_{\lambda}^\dagger a_{\lambda'}\rangle - U_\lambda U_{\lambda'} 
e^{-(\omega+\omega')t_0}\langle a_{\lambda} a_{\lambda'}\rangle\right]
\right.\nonumber\\
&& \qquad + \left[ \partial^j U_\lambda^* \partial_j 
U_{\lambda'} e^{-|\omega-\omega'|t_0}\langle a_{\lambda}^\dagger
a_{\lambda'}\rangle+\partial^j U_\lambda \partial_j U_{\lambda'}
e^{-(\omega+\omega')t_0}\langle a_{\lambda} a_{\lambda'}\rangle\right]
\nonumber\\
& &\qquad + \mu^2 \left.\left[ U_\lambda^* U_{\lambda'} 
e^{-|\omega-\omega'|t_0}\langle a_{\lambda}^\dagger a_{\lambda'}\rangle 
+  U_\lambda U_{\lambda'} e^{-(\omega+\omega')t_0}\langle a_{\lambda} 
a_{\lambda'}\rangle\right] \right\} \label{eq:DeltaRho}
\end{eqnarray}

We are seeking a lower bound on this quantity. It has been shown
\cite{F&RoUn} that
\begin{equation}
{\rm Re}\sum_{\lambda\lambda'} f(\lambda)^* f(\lambda')
e^{-|\omega-\omega'|t_0} \langle a_{\lambda}^\dagger a_{\lambda'}\rangle
\geq {\rm Re}\sum_{\lambda\lambda'} f(\lambda)^* f(\lambda')
e^{-(\omega+\omega')t_0}\langle a_{\lambda}^\dagger a_{\lambda'}\rangle.
\end{equation}
Upon substitution of this into Eq.~(\ref{eq:DeltaRho}) we have
\begin{eqnarray} 
\Delta\hat \rho & \geq & {\rm Re}\sum_{\lambda\lambda'}\left\{{\omega\omega'}
\left[ U_\lambda^* U_{\lambda'} \langle a_\lambda^\dagger
a_{\lambda'}\rangle - U_\lambda U_{\lambda'}\langle a_\lambda 
a_{\lambda'}\rangle\right]\right.\nonumber\\
& &\qquad +  \left[ \partial^j  U_\lambda^* \partial_j U_{\lambda'}
\langle a_\lambda^\dagger a_{\lambda'}\rangle+
\partial^j U_\lambda \partial_j U_{\lambda'}\langle a_\lambda 
a_{\lambda'}\rangle\right]\nonumber\\
& &\qquad + \mu^2\left. 
\left[ U_\lambda^* U_{\lambda'}\langle a_\lambda^\dagger
a_{\lambda'}\rangle +  U_\lambda U_{\lambda'}  \langle a_{\lambda} 
a_{\lambda'}\rangle\right]\right\} e^{-(\omega+\omega')t_0}\,.
                                          \label{eq:ineq1}
\end{eqnarray}  
We may now apply the inequalities proven in the Appendix. For the first and
third term of Eq.~(\ref{eq:ineq1}), apply Eq.~(\ref{eq:lowbound3}) with
$h_\lambda = \omega\, U_\lambda {\rm e}^{-\omega t_0}$ and
$h_\lambda = \mu\, U_\lambda {\rm e}^{-\omega t_0}$, respectively. For the 
second term of Eq.~(\ref{eq:ineq1}), apply Eq.~(\ref{eq:lowbound}) with
$A_{ij} = g_{ij}$ and 
$h_\lambda^i = \partial^i U_\lambda {\rm e}^{-\omega t_0}$. The result is
\begin{equation}
\Delta\hat \rho \geq -{1\over 2} \sum_\lambda \left(\omega_\lambda^2
U_\lambda^* U_{\lambda} + \partial^j U_\lambda^* \, \partial_j U_\lambda +
\mu^2 U_\lambda^* U_{\lambda}\right)e^{-2\omega_\lambda t_0}.
\end{equation}
This inequality may be re-written using the equation satisfied by the 
spatial mode functions:
\begin{equation}
\nabla^j\nabla_j U_\lambda +(\omega^2 -\mu^2)U_\lambda =0\,,
\end{equation}
to obtain
\begin{equation}
\Delta\hat \rho \geq - \sum_\lambda \left(\omega_\lambda^2
 + {1\over 4}\nabla^j\nabla_j\right) |U_\lambda|^2 {\rm e}^{-2
\omega_\lambda t_0}\, .                          \label{eq:qi2}
\end{equation}
Here $\nabla^i$ is the covariant derivative operator in the 
$t={\rm constant}$ hypersurfaces.
Therefore, given any metric which admits a global timelike Killing vector,
we can calculate the limitations on the negative energy
densities once the solutions to the wave equation in that curved background
are known. 

Note that although the local energy density may be more negative in
a given quantum state than in the vacuum, the total energy difference
integrated over all space is non-negative. This follows from the fact
that the normal-ordered Hamiltonian,
\begin{equation}
:H: \; = \int :T_{00}:\, \sqrt{-g}\, d^nx = \sum_\lambda \omega_\lambda \,
a_\lambda^\dagger a_\lambda \, ,
\end{equation}
is a positive-definite operator, so that $\langle :H: \rangle \geq 0$.

In the following sections, we will apply the general energy density 
inequality, Eq.~(\ref{eq:qi2}), to three and four-dimensional 
Robertson-Walker universes.

\section{Quantum Inequality in a 3D Closed Universe}

Let us consider the three-dimensional spacetime with a length element given
by 
\begin{equation}
ds^2 = - dt^2 + a^2 \left(d\theta^2 + \sin^2\theta d\varphi^2\right)
\label{eq:Sphere}
\end{equation}
Here constant time slices of this universe are two spheres of radius
$a$. The wave equation on this background with a coupling 
of strength $\xi$ to the Ricci scalar $R$ is
\begin{equation}
\BoxOp\phi -(\mu^2+\xi R)\phi = 0.
\end{equation}
For the metric Eq.~(\ref{eq:Sphere}), this equation becomes
\begin{equation}
-\partial_t^2\phi + {1\over{a^2\sin\theta}}\partial_\theta\left(\sin\theta
\partial_\theta\phi\right) + {1\over{a^2\sin^2\theta}}\partial_\varphi^2\phi
-[\mu^2 + \xi\left(2\over a^2\right)]\phi = 0,
\end{equation}
which has the solutions
\begin{eqnarray}
f_{lm}(t,\theta,\varphi) = {1\over{\sqrt{2a^2\omega}}} {\rm Y}_{lm}(\theta
,\varphi) e^{-i\omega_l t}& l = 0,1,2, \cdots\\
 & -l\leq m\leq l,
\end{eqnarray}
with eigenfrequencies
\begin{equation}
 \omega_l = a^{-1} \sqrt{ l(l+1) + 2\xi + \mu^2 a^2} \,.
\end{equation}
The ${\rm Y}_{lm}$ are the usual spherical harmonics with unit
normalization.  The coupling parameter $\xi$ here can be seen to contribute
to the wave functions as a term of the same form as the mass.
In this paper, however, we will look only at minimal coupling ($\xi = 0$).
The lower bound on the energy density is then given by  
\begin{equation}
\Delta\hat \rho \geq - {1\over {2a^2} }
\sum_{l= 0}^\infty \sum_{m=-l}^{+l} \omega_l |{\rm Y}_{lm}(\theta,\varphi)|^2 
e^{-2\omega_l t_0} - {1\over 8 a^2}\nabla^i \nabla_i
\sum_{l= 0}^\infty \sum_{m=-l}^{+l} {1\over \omega_l}|{\rm Y}_{lm}
(\theta,\varphi)|^2 e^{-2\omega_l t_0}
\label{eq:spherebound}
\end{equation}
However the spherical harmonics obey a sum rule \cite{Jackson}
\begin{equation}
\sum_{m=-l}^{+l} |{\rm Y}_{lm}(\theta,\varphi)|^2  = {{2l+1}\over{4\pi}}.
\end{equation}
We immediately see that the difference inequality is independent of
position, as expected from the spatial homogeneity, and that the second
term of Eq.~(\ref{eq:spherebound}) does not contribute.  We have
\begin{equation}
\Delta\hat \rho \geq - {1\over {8\pi a^2} }
\sum_{l= 0}^\infty  (2l+1)\, \omega_l  
e^{-2\omega_l t_0}. 
\end{equation}
This summation is finite due to the exponentially decaying term. We are
now left with evaluating the sum for a particular set
of values for the mass $\mu$, the radius $a$, and the sampling time $t_0$.  
Let $\eta = {t_0 / a}$.  Then
\begin{equation} 
\Delta\hat \rho \geq  -{1\over{16\pi t_0^3}}\left[2\eta^3\sum_{l=1}^\infty 
(2l+1)\, \tilde\omega_l \, 
    {\rm e}^{-2\eta\tilde\omega_l}\right]= -{1\over{16\pi t_0^3}}
F(\eta,\mu)                   \label{eq:qi3d}
\end{equation}
where
\begin{equation}\tilde\omega_l = \sqrt{l(l+1)+a^2 \mu^2}
\end{equation}
The coefficient $-{1/(16\pi t_0^3)}$ is the right hand side of the inequality
for the case of a massless field in an infinite 3-D Minkowski space.
The ``scale function'', $F(\eta,\mu)$, represents how the mass and the 
curvature of the closed spacetime affects the difference inequality.    

\subsection{Massless Case}
In terms of the variable $\eta = {t_0 / a}$,
which is the ratio of the sampling time to the radius of the universe, we can
write the above expression for $F(\eta,0)$ as
\begin{equation} 
F(\eta) = F(\eta,0) = 2\eta^3\sum_{l=1}^\infty 
(2l+1)\sqrt{l(l+1)}\, {\rm e}^{-2\eta\sqrt{l(l+1)}}
\end{equation}
A plot of $F(\eta)$ is shown in Figure 1.  In the limit of 
$\eta\rightarrow 0$, when the sampling time is very small or 
the radius of the universe has become so large that it approximates 
flat space, then the function $F(\eta)$ approaches 1, yielding
the flat space inequality. .

We can look at the inequality in the two asymptotic regimes of $\eta$. In the
large $\eta$ regime each term of greater $l$ in the exponent will decay away 
faster than the previous.  The $l = 1$ term yields a good approximation.
In the other regime, when the sampling time is small compared to the radius,
we can use the Plana summation formula to calculate the summation explicitly:
\begin{equation}
\sum_{n=1}^\infty f(n) + {1\over 2} f(0) = \int_0^\infty f(x)dx + i
\int_0^\infty {f(ix)-f(-ix) \over {{\rm e}^{2\pi x} -1}} dx \,,
                                              \label{eq:Plana}
\end{equation}
where
\begin{equation}
f(x) = (2x+1)\sqrt{x(x+1)}\, {\rm e}^{-2\eta\sqrt{x(x+1)}}.
\end{equation}
Immediately we see that for our summation $f(0) = 0$. The first integral
can be done with relative ease yielding
\begin{equation}
\int_0^\infty f(x) dx =
\int_0^\infty (2x+1)\sqrt{x(x+1)}\, 
             {\rm e}^{-2\eta\sqrt{x(x+1)}}dx = \frac{1}{2\eta^3} \,.
\end{equation}
This term reproduces the flat space inequality. The second integral 
in Eq.~(\ref{eq:Plana}) therefore contains all the corrections due to 
non-zero curvature  of the spacetime.  Since $\eta$ is small,
we can Taylor expand the exponent around $\eta = 0$, and then 
keep the lowest order terms.  One finds
\begin{equation}
i\int_0^\infty {f(ix)-f(-ix) \over {e^{2\pi x} -1}} dx \sim
- I_0 -\eta I_1 + \cdots \,,
                                     \label{eq:expans}
\end{equation}
where
\begin{equation}
I_0 = \int_0^\infty \sqrt{2x}\; {\left( 2x\sqrt{\sqrt{x^2+1}-x} +
\sqrt{\sqrt{x^2+1}+x}\right) \over {{\rm e}^{2\pi x} -1}}\; dx 
   \approx 0.265096
\end{equation}
and
\begin{equation}
I_1 = \int_0^\infty {4x(2x^2-1) \over {{\rm e}^{2\pi x} -1}} dx 
= -\frac{2}{15} \,.
\end{equation}

  Therefore the function $F(\eta)$ in the small $\eta$ limit is given by
\begin{equation}
F(\eta) \simeq 1 - 0.530192\, \eta^3 + {4 \over 15} \eta^4 + O(\eta^5)
         + \cdots
\end{equation}
and in the large $\eta$ limit by
\begin{equation}
F(\eta) \simeq 6\sqrt{2}\, \eta^3\, {\rm e}^{-2\sqrt{2}\eta} + \cdots \,.
                              \label{eq:large_eta}
\end{equation}
Both of these asymptotic forms are plotted along with the exact form of
$F(\eta)$ in Figure 1.  The graph shows that the two asymptotic limits
follow the exact graph to a very large precision except in the interval
of $ 1 < \eta <2$.  
These results for the function $F(\eta)$, combined with 
Eq.~(\ref{eq:qi3d}), yield the difference inequality for the massless
scalar field.

\begin{figure}
\epsfbox{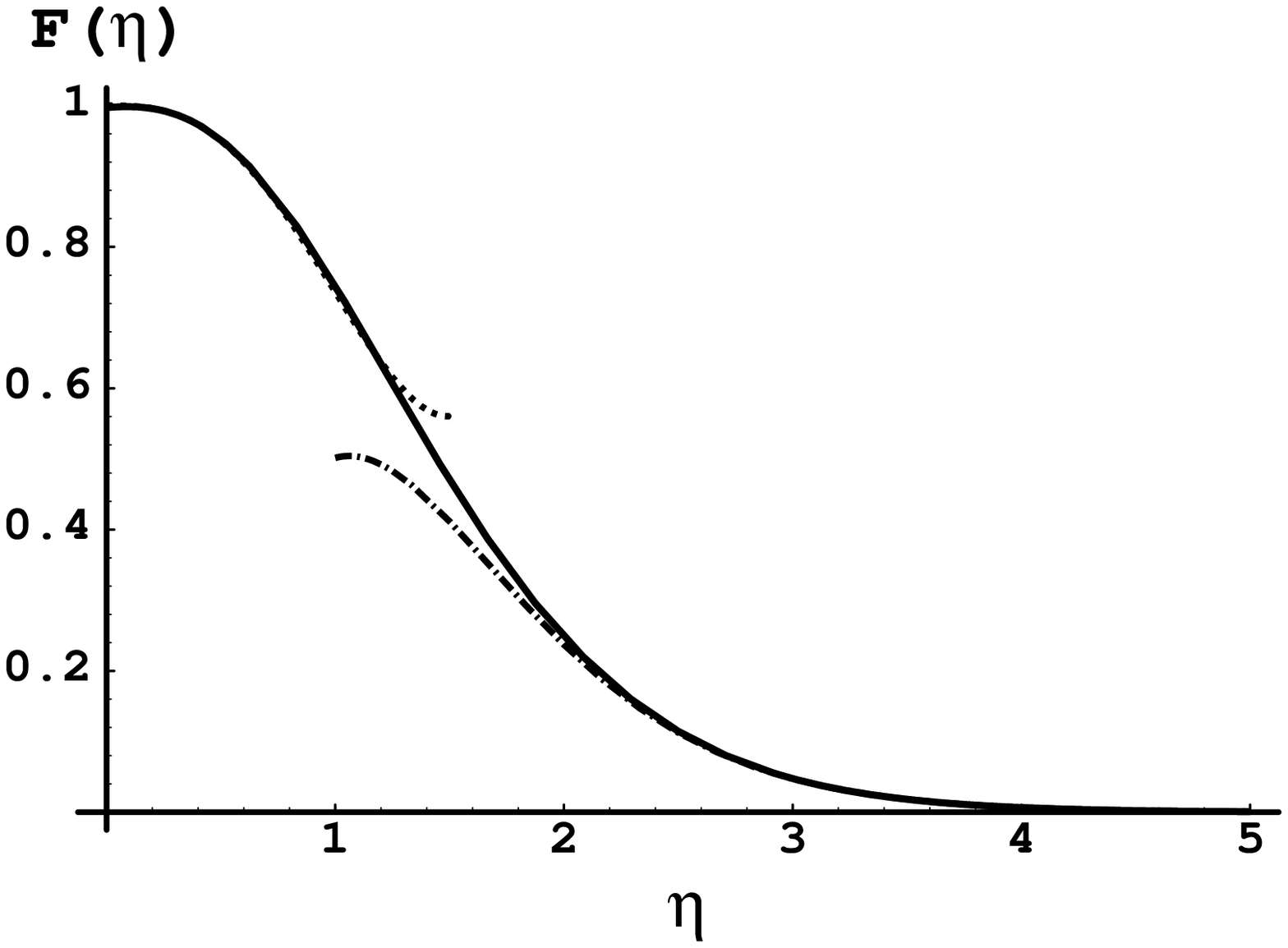}
\begin{caption}[]
  
Plot of the Scale Function $F(\eta)$ and its asymptotic forms.
The solid line is the exact form of the function.  The dotted
line is the small $\eta$ approximation, while the dot-dash 
line is the large $\eta$ approximation.
\end{caption}
\end{figure}

     The difference inequality does not require knowledge of
the actual value of the renormalized vacuum energy, $\rho_0$. However, 
if we wish to obtain a bound on the energy density itself, we must combine 
the difference inequality and $\rho_0$. One procedure for computing $\rho_0$
is analogous to that used to find the Casimir energy in flat spacetime
with boundaries: one defines a regularized energy density, subtracts the 
the corresponding flat space energy density, and then takes the limit in 
which the regulator is removed. A possible choice of regulator is to insert
a cutoff function, $g(\omega)$, in the mode sum and define the regularized
energy density as
\begin{equation}
\rho_{reg} =  \frac{1}{8 \pi a^2} \sum_\lambda \omega_\lambda \,
               g(\omega_\lambda) =
  \frac{1}{8 \pi a^2}  \sum_{l= 0}^\infty  (2l+1)\, \omega_l\, g(\omega_l)
        \,.
\end{equation}
In the limit that $a \rightarrow \infty$, we may replace the sum by an
integral and obtain the regularized flat space energy density:
\begin{equation}
\rho_{FS reg} =  \frac{1}{4 \pi}  \int_0^\infty {\rm d} \omega \, \omega\,
               g(\omega)        \,.
\end{equation}
The renormalized vacuum energy density may then be defined as
\begin{equation}
\rho_0 = \lim_{g \rightarrow 1} (\rho_{reg} - \rho_{FS reg}) \,.
\end{equation}
The vacuum energy density so obtained will be denoted by $\rho_{Casimir}$.
An analogous procedure was used in Ref. \cite{F75} to obtain the vacuum
energy density for the conformal scalar field in the 4D Einstein universe.
An explicit calculation for the present case, again using the Plana
summation formula, yields
\begin{equation}
\rho_{Casimir} =  - {I_0 \over{8\pi a^3}} 
               \approx - { 0.265096 \over{8\pi a^3}} .
\end{equation}
This agrees with the result obtained by Elizalde \cite{Elizalde} using the
zeta function technique. In this case $\rho_{Casimir} < 0$,
whereas the analogous calculation for the conformal scalar field in this
three-dimensional spacetime yields a positive vacuum energy density,
$\rho_{Casimir} = 1/{96\pi a^3}$. It should be noted that this procedure
works in this case only because  the divergent part of $\rho_{reg}$ is
independent of $a$. More generally, there may be curvature-dependent
divergences which must also be removed. 

Let us now return to the explicit forms of the difference inequality. In
the limit that $t_0 \gg a$, Eqs.~(\ref{eq:qi3d}) and (\ref{eq:large_eta})
yield
\begin{equation}
\Delta\hat \rho \geq 
-{{3\sqrt{2}}\over{8\pi a^3}}\, {\rm e}^{-2\sqrt{2}\, t_0/a}
    \,.                               
\end{equation}
Similarly, in the limit that $t_0 \ll a$, we find
\begin{equation}
\Delta\hat \rho \geq  -{1\over{16\pi t_0^3}} - \rho_{Casimir} -
{{t_0}\over{60\pi a^4}} + \cdots  \, . \label{eq:qi3}
\end{equation}
Thus the bound on the renormalized energy density in an arbitrary
quantum state in the latter limit becomes
\begin{equation}
\hat \rho_{ren} \geq  -{1\over{16\pi t_0^3}} - 
{{t_0}\over{60\pi a^4}} + \cdots  \, .  \label{eq:qi4}
\end{equation}

From either of Eqs.~(\ref{eq:qi3}) or (\ref{eq:qi4}) in
the $a \rightarrow \infty$ limit we obtain the flat space limit.
Here $\rho_0 = 0$ so $\Delta \rho = \rho$.
In 3D flat space we can write the quantum inequality in a more
covariant form as
\begin{equation}
\hat\rho = {\tau_0 \over \pi} \int_{-\infty}^\infty 
{\langle T_{\mu\nu}u^\mu u^\nu \rangle d\tau \over {\tau^2 + \tau_0^2}}
\geq -{1\over 16\pi \tau_0^3},
\end{equation}
In the other limit when the sampling time becomes long, we find
that $\Delta\hat \rho$ decays exponentially as a function of the sampling 
time. This simply reflects the fact that the difference in energy density
between an arbitrary state and the vacuum state satisfies the averaged
weak energy condition, Eq.~(\ref{eq:awec}).  

\section{4D Robertson-Walker Universes}

Now we will apply the same method to the case of the three homogeneous and
isotropic universes given by the four-dimensional static 
Robertson-Walker metrics.  Here we have 
\begin{equation}
[\epsilon = 0]: \qquad g_{ij}dx^i dx^j = dx^2 + dy^2 + dz^2
\end{equation}
for the flat universe with no curvature (Minkowski spacetime).  
For the closed universe with
constant radius $a$, i.e. the universe of constant positive curvature, the
spatial length element is given by.
\begin{equation}
[\epsilon = 1]: \qquad  g_{ij}dx^i dx^j = a^2 \left[ d\chi^2 + \sin^2\chi
(d\theta^2+\sin^2\theta d\varphi^2)\right] \,,  \label{eq:openmetric}
\end{equation}
where $0 \leq \chi < \pi$,\quad $0 \leq \theta < \pi$, and 
$ 0 \leq \varphi <2\pi$. 
The open universe $[\epsilon = -1]$ is given by making the replacement 
$\sin\chi \rightarrow \sinh\chi$ in Eq.~(\ref{eq:openmetric}) and now allowing
$\chi$ to take on the values $0 \leq \chi < \infty$.  To find the lower
bound of the quantum inequalities above we must solve for the eigenfunctions
of the covariant Helmholtz equation
\begin{equation}
\nabla_i \nabla^i U_\lambda({\bf x}) + (\omega^2_\lambda - \mu^2)
U_\lambda({\bf x}) = 0
\end{equation}
where $\mu$ is the mass of the scalar field and $\omega_\lambda$ is the
energy. A useful form of the solutions for this case is given by Parker
and Fulling \cite{Parker}. 

\subsection{Flat and Open Universes in 4D}

One finds that in the notation developed in Section 2, the 
spatial portion of the wave functions for flat (Euclidean) space 
is given by (continuum normalization)
\begin{eqnarray}
[\epsilon = 0]:& U_{\bf k}({\bf x}) = [2\omega (2\pi)^3]^{-1/2} 
e^{i{\bf k} \cdot {\bf x}}, \\
 & \omega = \sqrt{|\bf k|^2 + \mu^2}\\
 & {\bf k} = (k_1,k_2,k_3)\qquad (-\infty < k_j < \infty).\nonumber
\end{eqnarray}
It is evident that $|U_{\bf k}|^2$ will be independent
of position. This immediately removes the second term of the 
inequality in Eq.~(\ref{eq:qi2}).

In the open universe the spatial functions are given by
\begin{eqnarray}
[\epsilon = -1]:& U_{\lambda}({\bf x}) = (2a^3\omega_q)^{-1/2} 
 \Pi^{(-)}_{ql}(\chi) {\rm Y}_{lm}(\theta,\varphi), \\
 & \omega_q = \sqrt{{(q^2+1)\over a^2} + \mu^2},\\
 & \lambda = (q,l,m).\nonumber
\end{eqnarray}
Here, $0 <q<\infty;\quad  l = 0, 1, \cdots;$ and $m = -l, -l+1, \cdots, +l$.
The sum over all states involves an integral over the radial momentum $q$. 
The functions $\Pi^{(-)}_{ql}(\chi)$ are given in Equation 5.23 of 
Birrell and Davies\cite{Brl&Dv}. Apart from the normalization factor,
they are 
\begin{equation}
\Pi^{(-)}_{ql}(\chi) \propto \sinh^l\chi
\left( d\over d\cosh\chi\right)^{l+1} \cos q\chi.
\end{equation} 
As with the mode functions of the 3-dimensional closed spacetime above,
the mode functions of the open 4-dimensional universe satisfy an addition
theorem \cite{Parker,Lif&Khal,Bunch77}
\begin{equation}
\sum_{lm} |\Pi^{(-)}_{ql}(\chi) {\rm Y}_{lm}(\theta,\varphi)|^2 =
{q^2\over 2\pi^2}.
\end{equation}
Since the addition theorem removes any spatial dependence, we again get no
contribution from the Laplacian term of the quantum inequality,
Eq.~(\ref{eq:qi2}). Upon
substitution of the mode functions for both the flat and open universes
into the quantum inequality and using the addition theorem in the open
spacetime case we have
\begin{equation}
[\epsilon = 0]:\qquad\Delta\hat \rho \geq -{1\over 16 \pi^3}
\int_{-\infty}^\infty d^3k\, \omega_{\bf k}\, 
                                       {\rm e}^{-2\omega_{\bf k}t_0}
\end{equation}
and 
\begin{equation}
[\epsilon = -1]:\qquad\Delta\hat \rho \geq - {1\over 4\pi^2 a^3}
\int_0^\infty dq\, q^2\, \omega_q\, {\rm e}^{-2\omega_q t_0}, 
\end{equation}
respectively.  The 3-dimensional integral in momentum space can be carried
out by making a change to spherical momentum coordinates. The two cases can
be written compactly as
\begin{equation}
\Delta\hat\rho \geq -{1\over 4 \pi^2} \int_0^\infty dk\, k^2\, 
\tilde\omega\, {\rm e}^{-2\tilde\omega t_0}
\end{equation}
where 
\begin{equation}
\tilde\omega = \tilde\omega(k,a,\mu) = \left(k^2 - {\epsilon/ a^2} 
+\mu^2 \right)^{-1/2}\, .
\end{equation}
Note that $\epsilon = 0$ for flat space and $k = {q/ a}$
for the open universe.  This integral can be carried out explicitly in
terms of modified Bessel functions $K_n(z)$.  
The result is
\begin{equation}
\Delta\hat\rho \geq -{3\over 32 \pi^2 t_0^4} \left[ 
{1\over6}\left( z^3 K_3(z) - z^2 K_2(z)
\right) \right]  = -{3\over 32 \pi^2 t_0^4} G(z)  
\end{equation}
where
\begin{equation}
z = 2 t_0 \sqrt{  \mu^2 - {\epsilon\over a^2} }.
\end{equation}

The coefficient $ -3/ (32\pi^2 t_0^4)$ is the lower bound on $\hat \rho$
found in Refs. \cite{F&Ro95,F&RoUn} for a massless scalar field in 
Minkowski spacetime. The function $G(z)$ is the ``scale function'',
similar to that found in the case of the 3D closed universe. It is the
same function  found in Ref. \cite{F&RoUn} for Minkowski spacetime,
($\epsilon=0$) and is plotted in Figure 2.  Again we see that in the limit of
$z \rightarrow 0$ the scale function approaches unity, returning 
the flat space massless inequality in four dimensions \cite{F&Ro95}.  

\begin{figure}
\epsfbox{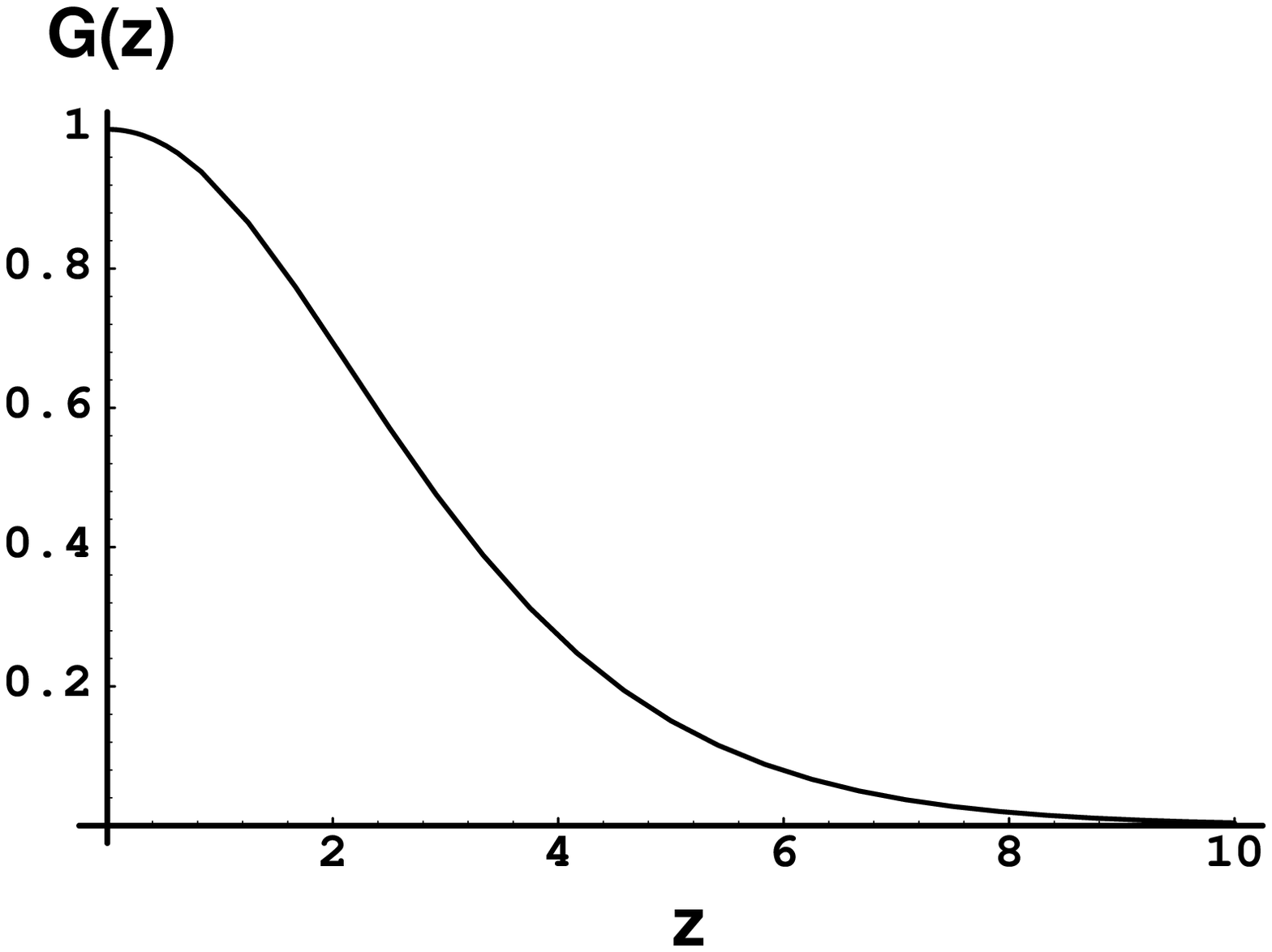}
\begin{caption}[]
  
Plot of the Scale Function $G(z)$ for the Open and Flat Universes.
\end{caption}
\end{figure}

\subsection{The Closed Universe}

In the case of the closed universe the spatial mode
functions are the 4-dimensional spherical harmonics, which have the form
\cite{Parker,Ford76} 
\begin{eqnarray}
[\epsilon = 1]:& U_{\lambda}({\bf x}) = (2\omega_n a^3)^{-1/2}\, 
 \Pi^{(+)}_{nl}(\chi)\, {\rm Y}_{lm}(\theta,\varphi), \\
 & \omega_n = \sqrt{{n(n+2)\over a^2} + \mu^2},\\
 & \lambda = (n,l,m).\nonumber
\end{eqnarray}
Here, $n = 0, 1, 2, \cdots;\quad  l = 0, 1, \cdots,n;$ and 
$m = -l, -l+1, \cdots, +l$.
The function  $\Pi^{(+)}$ is found from $\Pi^{(-)}$ by replacing
$\chi$ by $-i\chi$ and $q$ by $-i(n+1)$ \cite{Parker}.  Alternatively,
they can be written in terms of Gegenbauer polynomials \cite{Ford76,Erdelyi}
as
\begin{equation}
\Pi^{(+)}_{nl}(\chi) \propto \sin^l\chi \;{\rm C}^{l+1}_{n-l}(\cos \chi).
\end{equation}
In either case, the addition theorem is found from Eq.~11.4(3)
of Ref.\cite{Erdelyi} (with $p=2$ and $\xi = \eta$). This reduces to 
\begin{equation}
\sum_{lm} |\Pi^{(+)}_{nl}(\chi) {\rm Y}_{lm}(\theta,\varphi)|^2 
= {(n+1)^2\over 2\pi^2} \,,
\end{equation}
from which it is easy to show that the energy density inequality,
Eq.~(\ref{eq:qi2}), becomes
\begin{equation}
\Delta\hat \rho \geq - {1\over 4\pi^2 a^3}
\sum_{n=0}^\infty (n+1)^2\, \omega_n\, {\rm e}^{-2\omega_n t_0}. 
\end{equation}

If we use the variable $\eta = {t_0 / a}$ in the above
equation, we can simplify it to
\begin{equation}
\Delta\hat \rho \geq - {3\over 32\pi^2 t_0^4} H(\eta,\mu).
\end{equation}
Again we find the flat space solution in 4 dimensions, multiplied by
the scale function $H(\eta,\mu)$, which is defined as
\begin{equation}   
H(\eta,\mu) \equiv {8\over 3} \eta^4 \sum_{n=1}^\infty (n+1)^2 
\sqrt{n(n+2)+a^2\mu^2}\; {\rm e}^{-2\eta \sqrt{n(n+2)+a^2\mu^2}},
\end{equation}
and is plotted in Figure 3.  The scale function here
has a small bump in it, occurring at roughly $\eta \approx 0.5$ with a
height of 1.03245.  This may permit the magnitude of the negative energy
to be slightly greater for a massless scalar field in the Einstein universe
than is allowed in a flat universe for comparable sampling times.
A similar result was shown to exist for massive fields in a 2 
dimensional Minkowski spacetime \cite{F&RoUn}. 

\subsection{Massless asymptotic limits in the Einstein universe}

As with the 3-dimensional closed universe, we can find the 
asymptotic limits of this function. We again follow the method
of the previous section, assuming the scalar field is
massless, and making use of the Plana summation formula to
find
\begin{equation}
H(\eta,0) = {8\over 3}\eta^4\, \bigl( I_2 + I_3 \bigr) \, , 
\end{equation}
where
\begin{equation}
I_2 = \int_0^\infty (x+1)^2\sqrt{x(x+2)}
\, {\rm e}^{-2\eta\sqrt{x(x+2)}}\, dx
\end{equation}
and
\begin{eqnarray}
I_3 &=& 2 {\rm Re} \left[ i\int_0^\infty {(ix+1)^2\sqrt{ix(ix+2)}
{\rm e}^{-2\eta\sqrt{ix(ix+2)}}\over {\rm e}^{2\pi x} -1} \, dx \right] 
                                                        \nonumber \\
&=& \int_0^\infty \sqrt{2x}\: \frac{ (x^2-1)\sqrt{\sqrt{x^2+4}+x}
    -2x \sqrt{\sqrt{x^2+4}-x} }{{\rm e}^{2\pi x} -1} \:{\rm d}x \nonumber \\
 &\approx& -0.356109  \,.
\end{eqnarray}
The first integral can be done in terms 
of Struve, ${\bf H}_n(z)$, and Neumann, $N_n(z)$, functions, with the result
\begin{equation}
I_2 = {\pi\over 16} {d^2\over d\eta^2}\left[{1\over\eta}
\left( {\bf H}_1(2\eta) - N_1(2\eta)\right)\right] + \frac{4}{15} \eta \,.
                                   \label{eq:I2}
\end{equation}

If we follow the same procedure as in the previous section for defining
the renormalized vacuum energy density for the minimally coupled scalar
field in the Einstein universe, then we obtain
\begin{equation}
\rho_{Casimir} = {1\over{4\pi^2 a^4}}\, I_3 
       \approx - { 0.356109 \over{4\pi^2 a^4}} \,. \label{eq:4dCasimir}
\end{equation}
The same method yields $\rho_{Casimir} = 1/{480\pi^2 a^4}$ for the massless
conformal scalar field \cite{F75}. Here our result for the minimal
field, Eq.~(\ref{eq:4dCasimir}), differs from that obtained by Elizalde
\cite{Elizalde} using the zeta function method, $\rho'_{Casimir} = 
- 0.411502/{4\pi^2 a^4}$. This discrepancy probably reflects the fact that
the renormalized vacuum energy density is not uniquely defined. The 
renormalized stress tensor in a curved spacetime is only defined up to
additional finite renormalizations of the form of those required to
remove the infinities. In general this includes the geometrical tensors
$^{(1)}H_{\mu\nu}$ and $^{(2)}H_{\mu\nu}$. (See any of the references in
\cite{renorm} for the definitions of these tensors and a discussion of
their role in renormalization.) In the Einstein universe, both of
these tensors are nonzero and are proportional to $1/a^4$. Thus the
addition of these tensors to $\langle T_{\mu\nu} \rangle$ will change
the numerical coefficient in $\rho_{Casimir}$. The logarithmically
divergent parts of $\langle T_{\mu\nu} \rangle$ which are proportional
to $^{(1)}H_{\mu\nu}$ and $^{(2)}H_{\mu\nu}$ happen to vanish in the
Einstein universe, but not in a more general spacetime. In principle,
one should imagine that the renormalization procedure is performed in
an arbitrary spacetime, and only later does one specialize to a
specific metric. (Unfortunately, it is computationally impossible to do
this explicitly.) Thus the fact that a particular divergent term
happens to vanish in a particular spacetime does not preclude the 
presence of finite terms of the same form.

In the  small $\eta$ limit we can Taylor expand the Struve and Neumann 
functions in Eq.~(\ref{eq:I2}), and obtain
\begin{equation}   
H(\eta,0) = 1 + {1\over 3}\eta^2 + {(3+4\gamma)\over 12} \eta^4
+{8\over 3}(-0.356109)\eta^4 + {1\over 3}\eta^4 \log \eta 
   + O(\eta^6) + \cdots  ,
\end{equation}
where $\gamma$ is Euler's constant, which arises in the Taylor series
expansion of the Neumann function.  This is similar to that
of the 3-dimensional universe and again contains a term of the
form of the Casimir energy.  
In the large $\eta$ limit we again keep just the first term of
the series.  Both of the asymptotic forms are plotted with the
exact solution in Figure 3.  We see that the asymptotic form is
again a very good approximation except in the interval $1 < \eta
<2$, as was the case for the 3-dimensional closed universe.

\begin{figure}
\epsfbox{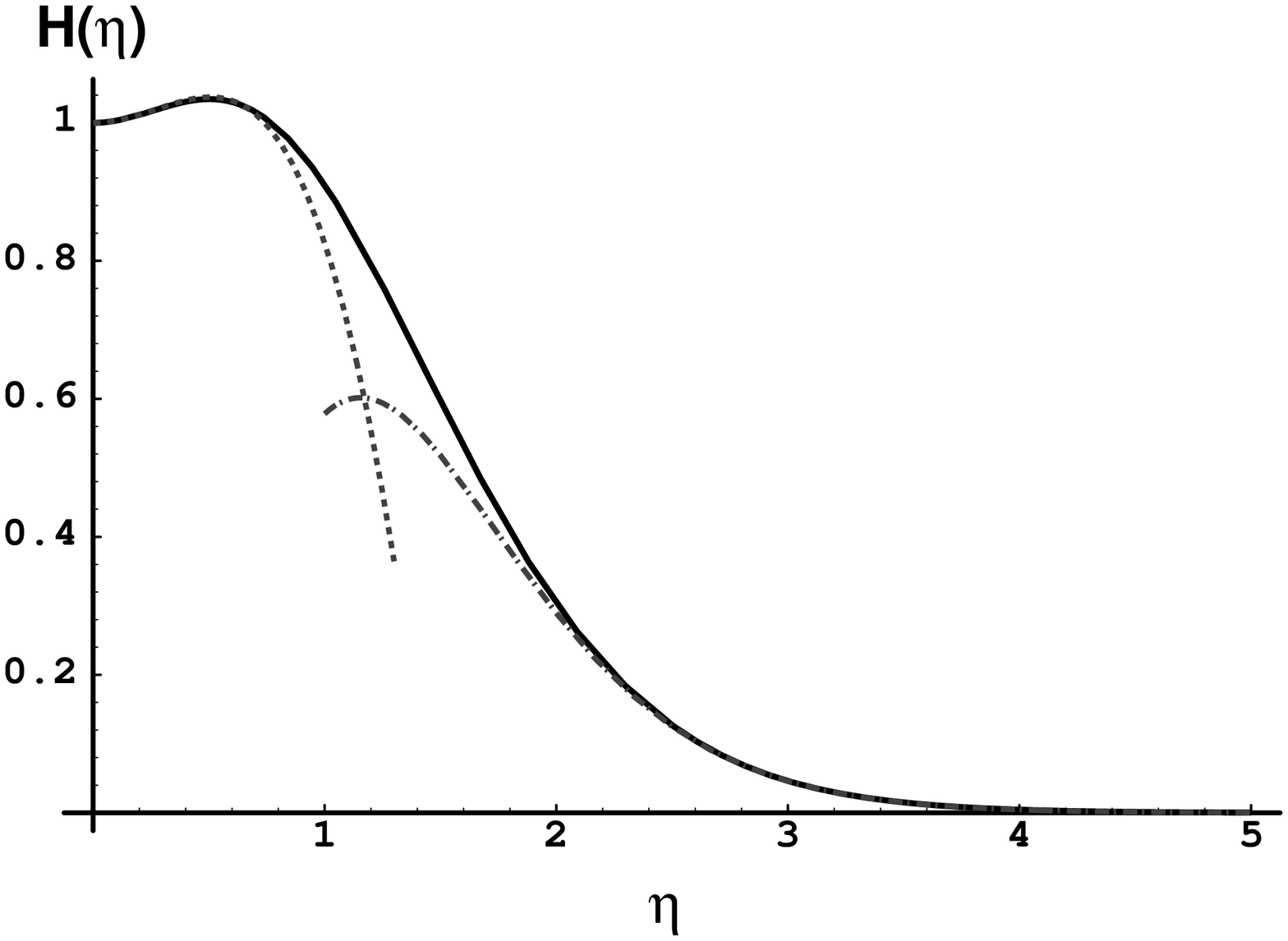}
\begin{caption}[]  

The Scale Function $H(\eta)$ for the Closed Universe.  The
solid line is the exact result.  The dotted line is the 
asymptotic expansion for small $\eta$,  while the dot-dash
curve is the large $\eta$ approximation.  The maximum occurs
at $\eta\approx 0.5$.
\end{caption}
\end{figure}

The difference inequalities for massless fields are then given by
\begin{equation}
\Delta\hat \rho \geq  -{3\over{32\pi^2 t_0^4}} \left[ 1 
 + {1\over 3}\eta^2 + {(3+4\gamma)\over 12} \eta^4
 + {1\over 3}\eta^4 \log \eta + \cdots \right] - \rho_{Casimir}
\end{equation}
for $t_0 \ll a$ and
\begin{equation}
\Delta\hat \rho \geq -{\sqrt{3}\over{\pi^2 a^4}}\, 
{\rm e}^{-2\sqrt{3}\, t_0/a} \qquad {\rm for}\quad t_0 \gg a\,.
\end{equation}
Using $\Delta\hat\rho =\hat\rho_{ren} - \hat\rho_{Casimir}$,
where $\hat\rho_{Casimir}$ is the expectation value of $\rho_{ren}$ in the
vacuum state, one can calculate the ``total,'' or more formally the 
renormalized energy density that would be constrained by the quantum
inequalities, subject to renormalization ambiguities. 
 For example in the 4-dimensional Einstein universe the
renormalized energy density inequality is constrained by
\begin{equation}
\hat \rho_{ren} \geq  -{3\over{32\pi^2 t_0^4}} \left[ 1 
 + {1\over 3}\eta^2 + {(3+4\gamma)\over 12} \eta^4
 + {1\over 3}\eta^4 \log \eta + O(\eta^6) + \cdots \right]
\end{equation}
for $t_0 \ll a$. Here the coefficient of the $\eta^4$ term would
be altered by a finite renormalization, but the rest of the expression 
is unambiguous. 
Similar expressions could be found for any of the other cases above.
In the case of the flat Robertson-Walker universe, there is no Casimir
vacuum energy.  Under such circumstances the difference inequality and
the renormalized energy density inequality are the same, and are free
of ambiguities.

\section{Discussion}

   In the previous sections, we have derived a general form for a
difference inequality for a massive, minimally coupled scalar field
in a static curved spacetime. This general inequality was then evaluated
explicitly in a 3-dimensional closed universe and in 4-dimensional
closed and open universes. The resulting inequalities include the effects
of the spacetime curvature. In the small sampling time limit, they 
reduce to the flat space forms plus curvature-dependent corrections.
These results lend support to the arguments given in Ref. \cite{F&Ro96}
to the effect that this should always be the case in curved spacetime.

An additional feature of these difference inequalities is that they
lead to averaged weak energy type conditions in the infinite sampling
time limit.  We found that in general
\begin{equation}
\Delta\hat \rho \; \equiv \; {t_0\over\pi}\int_{-\infty}^\infty 
{{\langle :T_{00}:\rangle dt} \over{t^2 + t_0^2}} \; \geq -{a_n \over t_0^n}
F_n(\eta)
\end{equation}
where $n$ is the dimensionality of the spacetime, $a_n$ is the coefficient
of the flat space inequality in $n$ dimensions, and $F_n(\eta)$ is the
scale function.  In all of the cases above the scale function decays
exponentially to zero in the large $\eta$ limit, which is the same as
the large $t_0$ limit.  In the $t_0 \rightarrow \infty$ limit the above 
expression reduces to
\begin{equation}
\int_{-\infty}^\infty \langle :T_{00}: \rangle dt 
= \int_{-\infty}^\infty \left[\langle\psi|T_{00}|\psi\rangle 
-\langle 0|T_{00}|0\rangle \right] dt \geq 0,
\end{equation}
where $|\psi\rangle$ is an arbitrary state.

The difference inequalities are derived directly from quantum field
theory and are in no way dependent upon the standard uncertainty relations.
It therefore appears that quantum field theory itself leads to constraints
on negative energy densities (fluxes) without any apriori assumptions.
In addition, the quantum inequalities in both flat space and these
generalized curved spacetimes are more restrictive on the magnitude and
duration of the allowable negative energy densities than is
the averaged weak energy condition.  Finally it
appears that the quantum inequalities themselves lead directly to AWEC type
conditions in many spacetimes.

\centerline{\bf Acknowledgements}

We would like to thank Thomas A. Roman for useful discussions and for
help with editing the manuscript.  This research was supported in part
by NFS Grant No. PHY-9507351.

\appendix
\section*{Appendix }
\label{sec:appendix}
\setcounter{equation}{0}
\renewcommand{\theequation}{A\arabic{equation}}

In this appendix, we wish to prove the following inequality: 
Let $A_{ij}$ be a real, symmetric $n \times n$ matrix with non-negative 
eigenvalues. (For the purposes of this paper, we may take either $n=2$,
for 3D spacetime, or $n=3$, for 4D spacetime.)
Further let $h_\lambda^i$ be a complex $n$-vector, which is also a
function of the mode label $\lambda$. Then in an arbitrary quantum
state $|\psi \rangle$, the inequality states that
\begin{equation}
{\rm Re}\sum_{\lambda,\lambda'} A_{ij} 
\left[ h^{i\,*}_\lambda h^j_{\lambda'} 
             \langle a_\lambda^\dagger a_{\lambda'} \rangle
\pm h^i_\lambda h^j_{\lambda'} \langle a_\lambda a_{\lambda'}\rangle \right]
\geq -{1\over 2}\sum_{\lambda} A_{ij} h^{i\,*}_\lambda h^j_\lambda \,.
\label{eq:lowbound}
\end{equation}
In order to prove this relation, we first note that 
\begin{equation}
A_{ij} = \sum_{\alpha =1}^n \kappa_\alpha V_i^{(\alpha)}\,V_j^{(\alpha)}\,,
\end{equation}
where the $V_i^{(\alpha)}$ are the eigenvectors of $A_{ij}$, and the
$\kappa_\alpha \geq 0$ are the corresponding eigenvalues. Now define the
hermitian vector operator
\begin{equation}
Q^i  =  \sum_{\lambda} \left( h^{i\,*}_\lambda a_\lambda^\dagger 
+ h^i_\lambda a_\lambda  \right) \,.
\end{equation}
Note that
\begin{equation}
\left\langle {Q^i}^\dagger A_{ij} Q^j \right\rangle =
\sum_{\alpha =1}^n \kappa_\alpha 
\left\langle Q^{i\,\dagger}  V_i^{(\alpha)}\,V_j^{(\alpha)} Q^j \right\rangle =
\sum_{\alpha =1}^n \kappa_\alpha ||V_i^{(\alpha)}\, Q^i |\psi\rangle ||^2
\geq 0 \,.
\end{equation}
Furthermore,
\begin{equation}
\left\langle Q^{i\, \dagger} A_{ij} Q^j \right\rangle =
2\,{\rm Re}\sum_{\lambda,\lambda'} A_{ij} 
\left[ h^{i\,*}_\lambda h^j_{\lambda'} \langle a_\lambda^\dagger 
a_{\lambda'}\rangle
+ h^i_\lambda h^j_{\lambda'} \langle a_\lambda a_{\lambda'}\rangle \right]
+ \sum_{\lambda} A_{ij} h^{i\, *}_\lambda h^j_\lambda \, ,
\end{equation}
from which Eq.~(\ref{eq:lowbound}) with the `$+$'-sign follows immediately.
The form of Eq.~(\ref{eq:lowbound}) with the `$-$'-sign can be obtained
by letting $h^j_\lambda \rightarrow i h^j_\lambda$.

As a special case, we may take $A_{ij} = \delta_{ij}$ and obtain
\begin{equation}
{\rm Re}\sum_{\lambda,\lambda'} 
\left[ {\bf h}_\lambda^*\cdot {\bf h}_{\lambda'} 
\langle a_\lambda^\dagger a_{\lambda'}\rangle
\pm {{\bf h}_\lambda}\cdot {\bf h}_{\lambda'}
     \langle a_\lambda a_{\lambda'}\rangle \right]
\geq -{1\over 2}\sum_{\lambda} |{\bf h}_\lambda|^2 \,.
\label{eq:lowbound2}
\end{equation}
As a further special case, we may take the vector ${\bf h}_\lambda$ to
have only one component, e.g. ${\bf h}_\lambda = (h_\lambda,0,0)$, in
which case we obtain
\begin{equation}
{\rm Re}\sum_{\lambda,\lambda'} 
\left[ {h_\lambda}^*  h_{\lambda'} 
\langle a_\lambda^\dagger a_{\lambda'}\rangle
\pm {h_\lambda} h_{\lambda'}
     \langle a_\lambda a_{\lambda'}\rangle \right]
\geq -{1\over 2}\sum_{\lambda} |h_\lambda|^2 \,.
\label{eq:lowbound3}
\end{equation}
This last inequality was originally proven in Ref. \cite{Ford91} for real
$h_\lambda$, and a simplified proof using the method adopted here is given
in Ref. \cite{F&RoUn}.


\begin{thebibliography}{99}

\bibitem{awec} Among the numerous papers on this topic are the following: 
F.J. Tipler, Phys. Rev. D {\bf 17}, 2521 (1978);
C. Chicone and P. Ehrlich, Manuscr. Math. {\bf 31}, 297 (1980);
G.J. Galloway, Manuscripta Math. {\bf 35}, 209 (1981);
A. Borde, Class. Quantum Grav. {\bf 4}, 343 (1987);
T.A. Roman, Phys. Rev. D {\bf 33}, 3526 (1986) and {\bf 37}, 546 (1988);
G. Klinkhammer, Phys. Rev. D {\bf 43}, 2542 (1991); and
R. Wald and U. Yurtsever, Phys. Rev D {\bf 44}, 403 (1991).

\bibitem{F&Ro95} L.H. Ford and Thomas A. Roman, Phys. Rev. D {\bf 51},
                 4277 (1995).

\bibitem{Yurtsever} U. Yurtsever, Phys. Rev. D {\bf 51}, 5797 (1995).

\bibitem{ford78} L.H. Ford, Proc. R. Soc. London {\bf A364}, 277 (1978)

\bibitem{F&Ro92} L.H. Ford and Thomas A. Roman, Phys. Rev. D {\bf 46},
                 1328 (1992).

\bibitem{F&Ro96} L.H. Ford and Thomas A. Roman, Phys. Rev. D {\bf 53},
                 5496 (1996). 

\bibitem{Mitch} Michael J. Pfenning and L.H. Ford, manuscript in preparation.

\bibitem{Alcu94} Miguel Alcubierre, Class. Quantum Grav. {\bf 11}, l73
                 (1994).

\bibitem{Brl&Dv} N.D. Birrell and P.C.W. Davies, 
                 {\it Quantum fields in curved space},
                 (Cambridge University Press, 1982).

\bibitem{F&RoUn} L.H. Ford and Thomas A. Roman,  ``Restrictions on
                 Negative Energy Density in Flat Spacetime'', Tufts
                 University preprint, TUTP-96-2, gr-qc/960703.

\bibitem{Jackson} See, for example, J.D. Jackson, 
                  {\it Classical Electrodynamics}, 2nd ed.
                 (John Wiley \& Sons, 1975), Eq.(3.69).

\bibitem{F75} L.H. Ford, Phys. Rev. D {\bf 12}, 2963 (1975).

\bibitem{Elizalde} E. Elizalde, J. Math. Phys. {\bf 35}, 3308 (1994).

\bibitem{Parker} Leonard Parker and S.A. Fulling, Phys. Rev. D {\bf 9},
                 341 (1974).

\bibitem{Lif&Khal} E.M. Lifshitz and I.M. Khalatnikov, Adv. Phys. {\bf 12},
                   185 (1963), Appendix J.
                   
\bibitem{Bunch77} T.S. Bunch, J. Phys. A {\bf 11}, 603 (1978).
                   
\bibitem{Ford76} L.H. Ford, Phys. Rev. D {\bf 14}, 3304 (1976).
                   
\bibitem{Erdelyi} A. Erdelyi, W. Magnus, F. Oberhettinger and F.G. Tricomi,
                  {\it Higher Transcendental Functions}, Vol. II,
                  (MacGraw-Hill, 1953). 

\bibitem{renorm} Ref. \cite{Brl&Dv}, Chap. 6; S.A. Fulling, {\it Aspects
of Field Theory in Curved Space-Time}, (Cambridge University Press, 1989),
Chap. 9; L.H. Ford, in {\it Cosmology and Gravitation}, M. Novello, ed.
(Editiones Fronti{\`e}res, 1994), Chap. 9. 

 \bibitem{Ford91} L.H. Ford, Phys. Rev. D {\bf 43}, 3972 (1991).
                 

\end{thebibliography}
\end{document}